\PassOptionsToPackage{dvipsnames}{xcolor}
\documentclass[sigconf, natbib=true, nonacm]{acmart}

\usepackage{acronym}
\usepackage{amsmath}
\usepackage[english]{babel}
\usepackage[shortlabels,inline]{enumitem}
\usepackage{graphicx}
\usepackage{hyperref}
\usepackage{cleveref}
\usepackage[utf8]{inputenc}
\usepackage{lipsum}
\usepackage{siunitx}
\usepackage[T1]{fontenc}
\usepackage{threeparttable}
\usepackage{csquotes}
\usepackage{bbm}
\usepackage{microtype}
\usepackage{dsfont}

\allowdisplaybreaks

\newcommand{\Figure}[1]{Fig.~\ref{#1}}

\newcommand{\Equation}[1]{\eqref{#1}}
\newcommand{\Equations}[2]{\eqref{#1} and~\eqref{#2}}

\newcommand{\Table}[1]{Table~\ref{#1}}

\newcommand{\Section}[1]{Section~\ref{#1}}

\acrodef{1D}[1-D]{one-dimensional}
\acrodef{2D}[2-D]{two-dimensional}
\acrodef{3D}[3-D]{three-dimensional}
\acrodef{ADMA}{Amplitude-Division Multiple Access}
\acrodef{ARE}{area rate efficiency}
\acrodef{AWGN}{Additive White Gaussian Noise}
\acrodef{BCA}{bicinchoninic assay}
\acrodef{BD}{basic threshold detection}
\acrodef{BER}{bit error rate}
\acrodef{BSC}{Binary Symmetric Channel}
\acrodef{CDF}{cumulative density function}
\acrodef{CDMA}{Code Division Multiple Access}
\acrodef{CIR}{channel impulse response}
\acrodef{CIRs}{channel impulse responses}
\acrodef{CSI}{channel state information}
\acrodef{CSK}{concentration shift keying}
\acrodef{CV}{column volume}
\acrodef{CVS}{cardiovascular system}
\acrodef{DD}{differential threshold detection}
\acrodef{EX}{eraser}
\acrodef{FCQD}{fluorescent carbon quantum dot}
\acrodef{FP}{fluorescent protein}
\acrodef{FPs}{Fluorescent Proteins}
\acrodef{GFP}{green fluorescent protein}
\acrodef{GFPD}{\textit{green fluorescent protein variant "Dreiklang"}}
\acrodef{IMAC}{immobilised metal chelate affinity chromatography}
\acrodef{IoBNT}{Internet of Bio-Nano-Things}
\acrodef{IPTG}{Isopropyl-$\beta$-D-thiogalactopyranosid}
\acrodef{IR}{impulse response}
\acrodef{ISI}{inter-symbol interference}
\acrodef{IUI}{inter-user interference}
\acrodef{lac operon}[]{lactose operon}
\acrodef{LB}{lysogeny broth}
\acrodef{LED}{light emitting diode}
\acrodef{MC}{molecular communication}
\acrodef{MCDMA}{Molecular Code Division Multiple Access}
\acrodef{MDMA}{Molecular Division Multiple Access}
\acrodef{MIMO}{multiple-input multiple-output}
\acrodef{ML}{maximum likelihood}
\acrodef{MLE}{maximum likelihood estimator}
\acrodef{MLSE}{maximum likelihood sequence estimator}
\acrodef{MTDMA}{Molecular Time Division Multiple Access}
\acrodef{NMSED}{normalized minimum squared Euclidean distance}
\acrodef{OD600}{optical density at 600 nm}
\acrodef{ODE}{ordinary differential equation}
\acrodef{OOK}{ON-OFF keying}
\acrodef{PBS}{particle-based simulation}
\acrodef{PDE}{partial differential equation}
\acrodef{PWM}{pulse width modulation}
\acrodef{RSFPs}{Reversable switchable fluorescent proteins}
\acrodef{RX}{receiver}
\acrodef{RXs}{receivers}
\acrodef{SDMA}{Space Division Multiple Access}
\acrodef{SDS-PAGE}{sodium dodecyl sulfate polyacrylamide gel electrophoresis}
\acrodef{SER}{symbol error rate}
\acrodef{SM}{signaling molecule}
\acrodef{SMC}{synthetic molecular communication}
\acrodef{SMs}{signaling molecules}
\acrodef{SNR}{signal-to-noise ratio}
\acrodef{SPION}{superparamagnetic iron oxide nanoparticle}
\acrodef{TDD}{targeted drug delivery}
\acrodef{TDMA}{Time Division Multiple Access}
\acrodef{TX}{transmitter}
\acrodef{TXs}{transmitters}
\acrodef{UCA}{uniform concentration assumption}
\acrodef{UV}{ultraviolet}
\acrodef{VE}{viterbi equalizer}
\acrodef{w.r.t.}{with respect to}
\acrodef{wlog}[w.l.o.g.]{without loss of generality}
\acrodef{YFP}{yellow fluorescent protein}
\acrodef{Sc}{scenario}
\acrodef{ICG}{Indocyanine green}

\setcopyright{none}
\acmConference[NANOCOM '25]{The 12th Annual ACM International Conference on Nanoscale Computing and Communication}{October 23--25, 2025}{Chengdu, China}
\acmBooktitle{The 12th Annual ACM International Conference on Nanoscale Computing and Communication (NANOCOM '25), October 23--25, 2025, Chengdu, China}

\begin{document}
\title{Closed-Loop Molecular Communication with Local and Global Degradation: Modeling and ISI Analysis}
\author{Lukas Brand, Fardad Vakilipoor, Sören Botsch, Timo Jakumeit, Sebastian Lotter, Robert Schober, and Maximilian Schäfer}
\affiliation{%
  \institution{Friedrich-Alexander-Universität Erlangen-Nürnberg, Erlangen, Germany} \city{}\country{}
}

\renewcommand{\shortauthors}{Brand et al.}

\thispagestyle{plain}
\pagestyle{plain}
\begin{abstract}
This paper presents a novel physics-based model for signal propagation in closed-loop \ac{MC} systems, which are particularly relevant for many envisioned biomedical applications, such as health monitoring or drug delivery within the closed-loop human \ac{CVS}. Compared to open-loop systems, which are mostly considered in \ac{MC}, closed-loop systems exhibit different characteristic effects influencing \ac{SM} propagation. One key phenomenon are the periodic \ac{SM} arrivals at the \ac{RX}, leading to various types of \ac{ISI} inherent to closed-loop system. 
To capture these characteristic effects, we propose an analytical model for the \ac{SM} propagation inside closed-loop systems. The model accounts for arbitrary spatio-temporal \ac{SM} release patterns at the \ac{TX}, and incorporates several environmental effects such as fluid flow, \ac{SM} diffusion, and \ac{SM} degradation. Moreover, to capture a wide range of practically relevant degradation and clearance mechanisms, the model includes both local removal (e.g., due to \ac{SM} absorption into organs) and global removal (e.g., due to chemical degradation) of \acp{SM}. 
The accuracy of the proposed model is validated with \ac{3D} \acp{PBS}. Moreover, we utilize the proposed model to develop a rigorous characterization of the various types of \ac{ISI} encountered in closed-loop \ac{MC} systems.

\end{abstract}
\acresetall



\setlength{\belowdisplayskip}{1pt}
\setlength{\belowdisplayshortskip}{1pt}
%
\maketitle

\acresetall
\section{Introduction}\label{sec:intro}
\Ac{MC} is a growing research field at the intersection of life sciences and engineering. One main direction of \ac{MC} research lies in the medical sector, aiming at the development of innovative diagnosis and therapeutic strategies for future healthcare systems, focusing on applications such as early disease detection or targeted treatment \cite{Chahibi2013MolecularCommunicationSystem,Akyildiz_Moving}.
In early disease detection, chemical signals that are potentially indicative of diseases, are sensed and processed inside the human body by synthetic nanodevices such as synthetic cells or nanorobots \cite{akyildiz2015internet}. In other future applications of \ac{MC}, in-body devices are envisioned to be equipped with capabilities to transmit, receive, and process molecular signals in order to communicate with biological entities or with each other to coordinate their actions during disease treatment.

In most disease detection applications, the \ac{MC} signal is a {\em natural signal}, i.e., it emanates from a natural biological \ac{TX}, e.g., diseased tissue, and nanodevices act as \acp{RX}. For communicating nanodevices, the \ac{MC} signal is a {\em synthetic signal}, i.e., engineered nanodevices act as \acp{TX} and \acp{RX}. Irrespective of the signal's origin, in both cases, signal propagation from the \ac{TX} to the \ac{RX} takes place inside the human body, mostly along the \ac{CVS}~\cite{Kianfar_Wireless}. Hence, it is of vital importance for medical applications of \ac{MC} to understand the characteristics of molecular signal propagation inside the \ac{CVS}.

Many studies in \ac{MC} consider the flow-based transport of mol\-e\-cules in cylindrical geometries as an approximation for molecular signal propagation in the \ac{CVS}. Most earlier works focus on the development of basic models \cite{Chahibi2013MolecularCommunicationSystem, Wicke2018, Mosayebi2019EarlyCancerDetection, Schaefer2021}, where molecule transport in a {\em single blood vessel} is commonly modeled as advection-diffusion process, and {\em vessel networks} are modeled as equivalent electrical networks \cite{Chahibi2013MolecularCommunicationSystem, Gomez2021MarkovModelFlow, Jakumeit2025MolecularSignalReception}.
Besides these theoretical works, also a growing number of experimental studies explore flow-dominated \ac{MC} in cylindrical structures (tubes), see \cite{Lotter2023testbedII} for a recent review.
However, most of the existing theoretical and experimental studies consider {\em open-loop} systems, i.e., there exist perfect {\em sources} and {\em sinks} for the carrier medium (and, hence, also for the \acp{SM}).
Consequently, the communication-theoretical tools, e.g., \ac{ISI} mitigation strategies, developed in these studies, have been validated only for open-loop \ac{MC} systems\footnote{Ref.~\cite{Gomez2021MarkovModelFlow} is an exception here, since it considers a closed-loop system. However, a steady-state molecule distribution is assumed for the analysis in \cite{Gomez2021MarkovModelFlow}, limiting its applicability to communication system design.}.

In contrast, in practical application environments of in-body \ac{MC} systems, the carrier medium, i.e., blood, is circulating in a closed-loop inside the \ac{CVS}. This indicates a systematic gap between most \ac{MC} systems investigated so far, and the envisioned \ac{CVS}-based real-world \ac{MC} systems.
Some recent experimental studies have considered closed-loop \ac{MC} systems \cite{tuccitto2017fluorescent, Schaefer2024ChorioallantoicMembraneModel, Brand2024ClosedLoopMolecular}, accumulating evidence that closed-loop systems exhibit characteristic properties, which necessitate specifically tailored communication system designs.
In particular, novel types of interference were identified experimentally in \cite{Brand2024ClosedLoopMolecular} that may occur in closed-loop systems but not in open-loop systems. However, while a first phenomenological model for particle distribution in closed-loop systems has been proposed in \cite{Schaefer2024ChorioallantoicMembraneModel}, these recent findings have not been complemented yet by a comprehensive theory for signal propagation in closed-loop \ac{MC} systems.
For example, the model in \cite{Schaefer2024ChorioallantoicMembraneModel} does not reveal the impact of physical system parameters such as vessel diameters on \ac{SM} propagation, and it does not account for the impact of molecule degradation. Such molecule degradation effects are crucial to consider, as they may arise either naturally in the \ac{CVS}, e.g., from the clearance of \acp{SM} by organs such as the liver or the kidneys, or in experimental systems because of the intentional (partial) removal or inactivation of \acp{SM} \cite{scherer2025closed}. In summary, theoretical tools for the analysis and design of closed-loop \ac{MC} systems do currently not exist. In this paper, we aim to lay the foundation for closing this research gap. In particular, we make the following contributions:
\begin{itemize}
    \item We propose a physics-based model for flow-based closed-loop \ac{MC} systems, along with a novel closed-form analytical expression for the \ac{SM} concentration. The proposed model accounts for the impact of diffusion and flow on the \acp{SM} propagating in closed-loop systems. Moreover, it allows for arbitrary release profiles, and incorporates the influence of various  biological (e.g., clearance by organs \cite{Schaefer2024ChorioallantoicMembraneModel}) and synthetic (e.g., the intentional inactivation \cite{Brand2024ClosedLoopMolecular}) \ac{SM} degradation effects by a general damping term.
    \item For the first time, we present a comprehensive theoretical characterization of different types of \ac{ISI}, which only occur in closed-loop \ac{MC} systems and were observed only experimentally before.
    \item Utilizing the proposed models, we analyze the impact of the different types of \ac{ISI} occurring in closed-loop systems on long transmission sequences and investigate how localized degradation effects can mitigate \ac{ISI}. In particular, we study how molecule degradation impacts whether or not the considered closed-loop system can be represented by an equivalent open-loop system.
\end{itemize}

The remainder of this paper is organized as follows.
In \Section{sec:system_model}, we introduce the considered closed-loop system model. 
In \Section{sec:description_and_solution}, we develop an analytical model and validate it with \acp{PBS}. In \Section{sec:isi}, we use the developed model to characterize different types of \ac{ISI}. In \Section{sec:evaluation}, we present simulation results and \Section{sec:conclusion} concludes the paper.

%
\section{System Model}\label{sec:system_model}
As previously mentioned, in many future \ac{MC} applications such as drug delivery or health monitoring, signals propagate along the \ac{CVS} from a \ac{TX} to an \ac{RX}. Inside the \ac{CVS}, \acp{SM} are transported mainly via advection and diffusion. Moreover, \acp{SM} may undergo degradation reactions, either gradually due to their irreversible absorption by vessel walls, e.g., in organs such as the liver, or intentionally, e.g., by inactivation\footnote{Additionally, reversible boundary interactions, such as the adsorption and desorption of \acp{SM} at the vessel wall, may play a role as well. This topic will be investigated in future work.}.
\begin{figure}
    \centering
    \vspace{-0.3cm}
    \includegraphics[scale=.65]{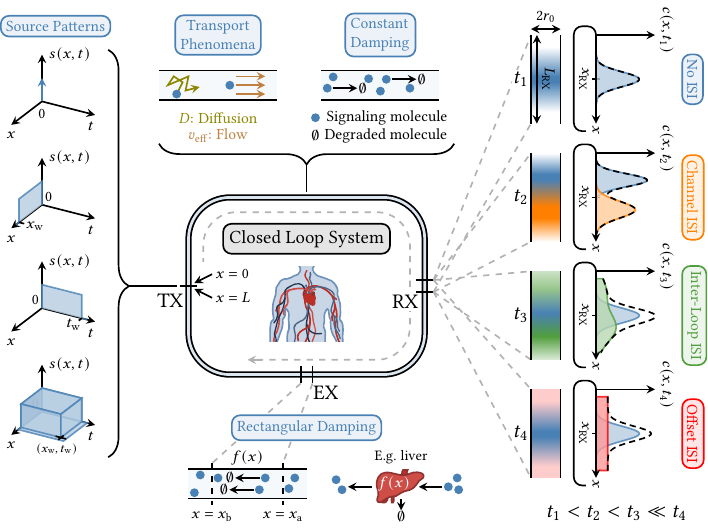}
    \vspace{-0.35cm}
    \caption{\small Schematic of the \ac{CVS} and the mimicking of its closed-loop characteristic as a \ac{1D} closed-loop system (center). \acp{SM} are transported mainly by diffusion and flow and are affected by global (e.g., \ac{SM} degradation) and spatially localized (e.g., \ac{SM} absorption into organs) degradation, modeled as constant and rectangular damping (top and bottom). At \ac{TX}, \acp{SM} can be injected with arbitrary temporal and spatial dynamics, modeled by the source term $s(x,t)$ (left). The different types of \ac{ISI} occurring in closed-loop systems are shown on the right hand side.}\Description{Schematic of the \ac{CVS} and the mimicking of its closed-loop characteristic as a \ac{1D} closed-loop system (center). \acp{SM} are transported mainly by diffusion and flow and are affected by global (e.g., \ac{SM} degradation) and spatially localized (e.g., \ac{SM} absorption into organs) degradation, modeled as constant and rectangular damping (top and bottom). At \ac{TX}, \acp{SM} can be injected with arbitrary temporal and spatial dynamics, modeled by the source term $s(x,t)$ (left). The different types of \ac{ISI} occurring in closed-loop systems are shown on the right hand side.}
    \label{fig:2:system model}
    \vspace{-0.6cm}
\end{figure}

The center of \Figure{fig:2:system model} shows a schematic of the \ac{CVS} as a highly complex closed-loop system. Modeling the \ac{CVS} in its full \ac{3D} geometry is very complex due to its intricate vascular network and the closed-loop topology. Therefore, we mimic the closed-loop characteristic of the \ac{CVS} by a \ac{3D} closed-loop pipe of constant radius $r_0$ and loop length $L$, i.e., a circumferential distance around the loop of length $L$. In this closed-loop pipe, \acp{SM} are transported by diffusion and laminar flow, and are affected by global or localized degradation reactions. Moreover, for high vascularization and due to the various phenomena affecting particle propagation, it has been shown in \cite{Schaefer2024ChorioallantoicMembraneModel} that the molecule transport in vascular networks is highly dispersive. Therefore, we assume that the \ac{SM} transport between the \ac{TX} and a transparent \ac{RX} occurs in the dispersive regime \cite{Jamali2019}, and we further simplify the \ac{3D} closed-loop system to a \ac{1D} closed-loop pipe. 
To this end, the \ac{SM} concentration $c(x,t)$ over time $t$ and space $x$ in the closed-loop pipe is governed by the following \ac{1D} advection-diffusion-reaction equation
{\setlength{\abovedisplayskip}{1pt}
 \setlength{\belowdisplayskip}{1pt}
\begin{align}
    \begin{split}
        \partial_t c(x,t) = D_{\mathrm{eff}} \partial_x^2 c(x,t) - v_{\mathrm{eff}} \partial_x c(x,t) - f(x)c(x,t) + s(x,t) \;,
    \end{split}
    \label{eq:PDE_general}
\end{align}}
where $ \partial_t $, $ \partial_x $, and $ \partial_x^2 $ denote the first order partial derivative with respect to time $ t $ and the first- and second-order partial derivatives with respect to space $ x $, respectively. Parameters $ D_{\mathrm{eff}} $ in $\si{\square\meter\per\second}$ and $ v_{\mathrm{eff}} $ in $\si{\meter\per\second}$ denote the effective diffusion coefficient of the \acp{SM} and the effective velocity of the fluid, respectively. The effective diffusion coefficient is given by $D_{\mathrm{eff}} = D \left( 1 + \frac{1}{48} \left( \frac{r_0 v_{\mathrm{eff}}}{D} \right)^2 \right)$~\cite[Eq.~(12)]{Wicke2018}, where $D$ is the diffusion coefficient in the respective \ac{3D} pipe of radius $r_0$. 
The terms $ f(x) $ and $ s(x,t) $ in \eqref{eq:PDE_general} are the spatially varying degradation function and the source term modeling \ac{SM} injection, respectively. In the following sections, the term $f(x)$ is used to model different types of localized and global effects affecting the \acp{SM} such as intentional or non-intentional degradation and absorption of \acp{SM} in specific parts of the system. As all of these effects have in common that they lead to fewer received molecules at the \ac{RX}, we refer to $f(x)$ as a damping term of $c(x,t)$. 
To account for the closed-loop system, \eqref{eq:PDE_general} is defined on spatial domain $ x \in [0, L] $, with periodic boundary conditions $\partial_x^k c(0, t) = \partial_x^k c(L, t)$, $\forall t \in \mathbb{R}^{+}, \ \forall k \in \mathbb{N}$, and initial condition $c(x,0) = 0$, where $\mathbb{R}^{+}$ and $\mathbb{N}$ denote the set of non-negative real numbers and the set of natural numbers, respectively.
%
\section{System Response}\label{sec:description_and_solution}
In this section, we propose an analytical model for \ac{SM} propagation in closed-loop systems by solving the \ac{PDE} in~\eqref{eq:PDE_general}. 
Due to the periodicity of the spatial domain, concentration $c(x,t)$ in \eqref{eq:PDE_general} can be expressed in terms of a Fourier series as
{\setlength{\abovedisplayskip}{2pt}
 \setlength{\belowdisplayskip}{2pt}
\begin{align}
    c(x,t) = \lim_{N\to\infty} \sum_{n=-N}^{N} \hat{c}_n(t) \mathrm{e}^{\mathrm{j} k_n x} \;,
    \label{eq:fourier_series}
\end{align}}
with Fourier coefficients $\hat{c}_n(t)$ and Fourier basis functions $\mathrm{e}^{\mathrm{j}k_n x}$. Here, $k_n \in \mathbb{R}$ and $j$ denotes the imaginary unit. We note that, for an exact solution, the number of terms in the sum in \eqref{eq:fourier_series} needs to be infinity, i.e., $ N \to \infty $. However, for mathematical tractability, in the following, we consider $N$ to be finite and large enough to well approximate $c(x,t)$, which we validate with \acp{PBS}, cf. \Section{sec:analytical_vs_PBS}.

The periodic boundary condition requires for the basis functions $ \mathrm{e}^{\mathrm{j} k_n L} = 1 $, yielding wavenumbers $ k_n = \frac{2\pi n}{L} $, $ n \in \mathbb{Z} $. Similarly, the degradation function $ f(x) $ and the source term $ s(x,t) $ in \eqref{eq:PDE_general} can be expanded into Fourier series as follows
{\setlength{\abovedisplayskip}{1pt}
 \setlength{\belowdisplayskip}{1pt}
\begin{align}\label{eq:f_and_s}
    f(x) = \sum_{m=-N}^{N} \hat{f}_m \mathrm{e}^{\mathrm{j} k_m x} \;, \quad s(x,t) = \sum_{n=-N}^{N} \hat{s}_n(t) \mathrm{e}^{\mathrm{j} k_n x}  \;,
\end{align}}
{\setlength{\abovedisplayskip}{1pt}
 \setlength{\belowdisplayskip}{1pt}
\begin{align}\label{eq:f_hat_and_s_hat}
    \hat{f}_m = \frac{1}{L} \int_0^L\!\! f(x) \mathrm{e}^{-\mathrm{j} k_m x} \, \mathrm{d}x \;, \quad \hat{s}_n(t) = \frac{1}{L} \int_0^L\!\! s(x,t) \mathrm{e}^{-\mathrm{j} k_n x} \, \mathrm{d}x \;,
\end{align}}
where $\hat{f}_m$ and $\hat{s}_n(t)$ are the Fourier coefficients for the expansion of $f(x)$ and $s(x,t)$, respectively.
Substituting~\eqref{eq:fourier_series} and ~\eqref{eq:f_and_s} into~\eqref{eq:PDE_general} yields the following system of \acp{ODE} for the $2N+1$ Fourier coefficients $ \hat{c}_n(t) $:
{\setlength{\abovedisplayskip}{1pt}
 \setlength{\belowdisplayskip}{1pt}
\begin{align}\label{eq:ODE}
    \!\partial_t \hat{c}_n(t) =& \!-D_{\mathrm{eff}} k_n^2 \hat{c}_n(t) \!- v_{\mathrm{eff}} \mathrm{j} k_n \hat{c}_n(t)  \!-\!\!\!\!\! \sum_{m=-N}^{N}\!\!\! \hat{f}_{n-m} \hat{c}_m(t) + \hat{s}_n(t) .\\[-0.7cm]
    \nonumber
\end{align}}
\subsection{Solution in Form of Matrix Exponential}
To solve the system of \acp{ODE} in \Equation{eq:ODE}, we first define a $(2N+1)\times 1$ vector of Fourier coefficients $\mathbf{\hat{c}}(t) = [\hat{c}_{-N}(t), \dots, \hat{c}_0(t), \dots, \hat{c}_N(t)]^\intercal$, where $[\cdot]^\intercal$ denotes transposition. Then, the system in \Equation{eq:ODE} can be rewritten as
{\setlength{\abovedisplayskip}{1pt}
 \setlength{\belowdisplayskip}{1pt}
\begin{align}\label{eq:expODE}
    \partial_t\hat{\mathbf{c}}(t) = \mathbf{A} \hat{\mathbf{c}}(t) + \hat{\mathbf{s}}(t) \;,
\end{align}}
where $\mathbf{A}$ is a $(2N+1)\times (2N+1)$ matrix and $\mathbf{\hat{s}}(t) = [\hat{s}_{-N}(t), \dots, \hat{s}_N(t)]^\intercal$ is the vector of coefficients $\hat{s}_{n}(t)$. 
The individual entries of matrix $\mathbf{A}$, $A_{n,m}$, are obtained by reformulating the system of \acp{ODE} in \eqref{eq:ODE} into vector form as follows
{\setlength{\abovedisplayskip}{1pt}
 \setlength{\belowdisplayskip}{1pt}
\begin{align}\label{eq:A_nm}
    A_{n,m} = \begin{cases} 
        -D_{\mathrm{eff}} k_n^2 - v_{\mathrm{eff}} \mathrm{j} k_n - \hat{f}_0 & \text{if } n = m \;, \\
        -\hat{f}_{n-m} & \text{if } n \neq m \;.
    \end{cases}
\end{align}}
The initial conditions for $\mathbf{\hat{c}}(t)$ at $t = 0$ in \eqref{eq:expODE} follow from the initial condition $c(x,0) = 0$ for \ac{PDE} \eqref{eq:PDE_general} as $\mathbf{\hat{c}}(0) = \mathbf{0}$, where $\mathbf{0} = [0, \dots, 0]^\intercal$ is a vector of zeros of length $(2N+1)\times 1$. Inserting $\mathbf{\hat{c}}(0) = \mathbf{0}$ into~\eqref{eq:expODE}, a solution for $\mathbf{\hat{c}}(t)$ can be obtained in terms of a matrix exponential as follows
{\setlength{\abovedisplayskip}{1pt}
 \setlength{\belowdisplayskip}{1pt}
\begin{align}\label{eq:int_hatc}
    \mathbf{\hat{c}}(t) = \int_0^t \mathrm{e}^{\mathbf{A} (t - \tau)} \mathbf{\hat{s}}(\tau) \, \mathrm{d}\tau \;.
\end{align}}
Inserting~\eqref{eq:int_hatc} into Fourier series \eqref{eq:fourier_series}, the concentration $c(x,t)$ can be expressed as
{\setlength{\abovedisplayskip}{1pt}
 \setlength{\belowdisplayskip}{1pt}
\begin{align}\label{eq:int_c}
\begin{split}
    c(x,t) &= \mathbf{e}^\intercal(x) \mathbf{\hat{c}}(t) = \mathbf{e}^\intercal(x) \int_0^t \mathrm{e}^{\mathbf{A} (t - \tau)} \mathbf{\hat{s}}(\tau) \, \mathrm{d}\tau \;,
\end{split}
\end{align}}
where the sum in \eqref{eq:fourier_series} is reformulated into a matrix vector multiplication using the $(2N+1)\times 1$ vector $\mathbf{e}(x) = [\mathrm{e}^{\mathrm{j} k_{-N} x}, \dots, \mathrm{e}^{\mathrm{j} k_N x}]^\intercal$ of Fourier basis functions.
\subsection{Global and Localized Damping}
The function $f(x)$ captures global and spatially localized degradation mechanisms such as a constant baseline degradation (e.g., spontaneous switching of \ac{GFPD}, a variant of \ac{GFP}~\cite{brakemann2011reversibly}, if used as \acp{SM}), and the localized clearance of \ac{SM} (e.g., \ac{ICG} clearance from the vascular system and accumulation in an organ, e.g., the liver~\cite{vakilipoor2025cam}). In the following, we first introduce a two-level damping function to model these effects, and then discuss several special cases.

\subsubsection{Two-Level Damping}\label{sec:TwoLVL}
To model spatially varying damping effects, we define $f(x)$ as a function with two distinct levels, representing a spatially localized region of higher damping within a global degradation environment\footnote{For simplicity, we consider a single high-damping region, but $f(x)$ can be readily extended to multiple regions with varying damping.}
 \setlength{\belowdisplayskip}{1pt}
{\setlength{\abovedisplayskip}{1pt}
\begin{align}\label{eq:ftwolevel}
    f(x) = \begin{cases} 
        \alpha & \text{if } x \in [x_{\mathrm{a}}, x_{\mathrm{b}}]  \;, \\
        \beta & \text{else} \;,
    \end{cases}
\end{align}}
where $[x_{\mathrm{a}}, x_{\mathrm{b}}]$ indicates the spatial interval with higher damping constant $\alpha\geq\beta$ in $\si{\per\second}$, $0 \leq x_{\mathrm{a}} < x_{\mathrm{b}} \leq L$, and $\beta$ in $\si{\per\second}$ reflects the baseline degradation rate. 
In order to obtain a solution for concentration $c(x,t)$ in~\eqref{eq:PDE_general}, the entries of matrix $\mathbf{A}$ in \eqref{eq:A_nm} need to be derived. Inserting \eqref{eq:ftwolevel} into \eqref{eq:f_hat_and_s_hat} the Fourier coefficients $\hat{f}_m$ can be obtained as
{\setlength{\abovedisplayskip}{1pt}
 \setlength{\belowdisplayskip}{1pt}
\begin{align}
    \hat{f}_m &= \frac{1}{L} \left( \int_0^{x_{\mathrm{a}}} \beta \mathrm{e}^{-\mathrm{j} k_m x} \, \mathrm{d}x + \int_{x_{\mathrm{a}}}^{x_{\mathrm{b}}} \alpha \mathrm{e}^{-\mathrm{j} k_m x} \, \mathrm{d}x + \int_{x_{\mathrm{b}}}^L \beta \mathrm{e}^{-\mathrm{j} k_m x} \, \mathrm{d}x \right) \nonumber \\
    & = \begin{cases} 
        \beta + \frac{(\alpha - \beta)(x_{\mathrm{b}} - x_{\mathrm{a}})}{L} & \text{if } m = 0  \;, \\
        \frac{(\alpha - \beta)}{\pi m} \mathrm{e}^{-\frac{\mathrm{j} k_m (x_{\mathrm{a}} + x_{\mathrm{b}})}{2}} \sin\left( \frac{k_m (x_{\mathrm{b}} - x_{\mathrm{a}})}{2} \right) & \text{if } m \neq 0  \;.
        \end{cases}\label{eq:f_hat_m}
\end{align}}
Inserting \eqref{eq:f_hat_m} into~\eqref{eq:A_nm}, the entries $A_{n,m}$ of the matrix $\mathbf{A}$ follow as
{\setlength{\abovedisplayskip}{1pt}
 \setlength{\belowdisplayskip}{1pt}
\begin{align}
    \!A_{n,m} \!&= \!\begin{cases} 
        -D_{\mathrm{eff}} k_n^2 - v_{\mathrm{eff}} \mathrm{j} k_n - \beta - \frac{(\alpha - \beta)(x_{\mathrm{b}} - x_{\mathrm{a}})}{L} & \! \! \text{if } n = m  , \\
        -\frac{(\alpha - \beta)}{\pi (n - m)} \mathrm{e}^{-\frac{\mathrm{j} k_{n-m} (x_{\mathrm{a}} + x_{\mathrm{b}})}{2}} \sin\left( \frac{k_{n-m} (x_{\mathrm{b}} - x_{\mathrm{a}})}{2} \right) & \! \! \text{if } n \neq m  .
    \end{cases}\!\!\label{eq:A_twolevel}
\end{align}}
Inserting matrix $\mathbf{A}$ with the entries in \eqref{eq:A_twolevel} into \eqref{eq:int_c} yields an analytical solution for the \ac{SM} concentration in a \ac{1D} closed-loop system accounting for the combined effects of diffusion, advection, and spatially varying damping.
\subsubsection{Special Cases of Damping}
The general solution for the entries of matrix $\mathbf{A}$ in \eqref{eq:A_twolevel} accounts for both global and localized damping effects simultaneously. By varying parameters $\alpha$ and $\beta$ in \eqref{eq:ftwolevel}, different special cases can be modeled, for which the entries $A_{n,m}$ simplify.

For \textit{localized damping}, $ \beta = 0 $, i.e., damping occurs only within $ [x_{\mathrm{a}}, x_{\mathrm{b}}] $ with rate $ \alpha > 0 $. This models scenarios, where damping is confined to a specific region, such as the absorption of \acp{SM} by organs, with negligible loss elsewhere.

For \textit{global damping}, $ \alpha = \beta > 0 $, resulting in $ f(x) = \beta $ across the entire spatial domain. This represents a system with homogeneous degradation, which models, e.g., a constant enzymatic degradation of \acp{SM} everywhere.

For \textit{no damping}, $ \alpha = \beta = 0 $, so $ f(x) = 0 $, modeling a system without degradation or damping effects. This simplifies \eqref{eq:PDE_general} to an advection-diffusion equation with known solutions \cite[Eq. (3)]{Schaefer2024ChorioallantoicMembraneModel}.
\subsection{Forms of Signaling Molecule Release}
In general, the function $s(x,t)$ can model any spatio-temporal release pattern of \acp{SM}. In the following, we derive the source term Fourier coefficients $\hat{\mathbf{s}}(t)$ needed for the solution in \eqref{eq:int_c} for two specific cases. The first case is an instantaneous release of \acp{SM} at a single point in space. The second case is a spatially extended distribution of \acp{SM} that is released instantaneously.

\subsubsection{\ac{SM} Point Release}\label{sec:dtdx}
For an instantaneous release of $N_{\mathrm{P}}$ \acp{SM} at $x = 0$ and $t = 0$, the source function and its Fourier coefficient vector (obtained by inserting the source function into \eqref{eq:f_hat_and_s_hat}) are given as
{\setlength{\abovedisplayskip}{1pt}
 \setlength{\belowdisplayskip}{1pt}
\begin{align}
    s(x,t) = N_{\mathrm{P}} \delta(x) \delta(t) \;, \quad \hat{\mathbf{s}}(t) = \frac{N_{\mathrm{P}}}{L} \delta(t) \mathbf{1} \;,\label{eq:impulse_release_source}
\end{align}}
where $\mathbf{1} = [1, \dots, 1]^\intercal$. 

Substituting $\hat{\mathbf{s}}(t)$ into~\eqref{eq:int_c}, $c(x,t)$ follows as
{\setlength{\abovedisplayskip}{1pt}
 \setlength{\belowdisplayskip}{1pt}
\begin{align}\label{eq:c_impulse}
    c(x,t) = \frac{N_{\mathrm{P}}}{L} \mathbf{e}^\intercal(x) \mathrm{e}^{\mathbf{A} t} \mathbf{1} \;.
\end{align}}
\subsubsection{Spatially Distributed \ac{SM} Release} \label{sssec:spatialRelease}
Considering $N_{\mathrm{P}}$ \acp{SM} released instantaneously at $t = 0$ spatially over $[0, x_{\mathrm{w}}]$, with $x_{\mathrm{w}} < L$,  the source function follows as
{\setlength{\abovedisplayskip}{1pt}
 \setlength{\belowdisplayskip}{1pt}
\begin{align}
    s(x,t) = N_{\mathrm{P}} \frac{\mathds{1}_{[0, x_{\mathrm{w}}]}(x)}{x_{\mathrm{w}}} \delta(t) \;, \label{eq:distributed_source}
\end{align}}
with indicator function $\mathds{1}_{[0, x_{\mathrm{w}}]}(x) = 1$ for $0 \leq x \leq x_{\mathrm{w}}$, and $0$ otherwise. Inserting \eqref{eq:distributed_source} into \eqref{eq:f_hat_and_s_hat} yields the source term's Fourier coefficients as
{\setlength{\abovedisplayskip}{1pt}
 \setlength{\belowdisplayskip}{1pt}
\begin{align}
    \hat{s}_n(t)\! &=\! \!\frac{N_{\mathrm{P}}}{L} \!\!\!\int_0^L\!\!\!\! \delta (t) \frac{\mathds{1}_{[0, x_{\mathrm{w}}]}(x)}{x_{\mathrm{w}}} \mathrm{e}^{-\mathrm{j} k_n x} \, \mathrm{d}x\! = \!\frac{N_{\mathrm{P}}}{L} \frac{\delta (t)}{x_{\mathrm{w}}}\!\!\int_0^{x_{\mathrm{w}}}\!\!\!\! \mathrm{e}^{-\mathrm{j} k_n x} \!\, \mathrm{d}x ,
\end{align}}
and the source's Fourier coefficient vector $\mathbf{\hat{s}}(t)$ follows as 
{\setlength{\abovedisplayskip}{1pt}
 \setlength{\belowdisplayskip}{1pt}
\begin{align}\label{eq:source_spacial_distribution}
    \mathbf{\hat{s}}(t) = \frac{N_{\mathrm{P}}}{L} \delta(t) \boldsymbol{\phi} \;, \quad \phi_n = \begin{cases} 
     \frac{1}{x_{\mathrm{w}}}\frac{1 - \mathrm{e}^{-\mathrm{j} k_n x_{\mathrm{w}}}}{\mathrm{j} k_n} & n \neq 0 \;, \\
     1& n = 0 \;,
     \end{cases}
\end{align}}
where $\boldsymbol{\phi} = [\phi_{-N}, \dots, \phi_N]^\intercal$.
Inserting \Equation{eq:source_spacial_distribution} in \eqref{eq:int_c}, $c(x,t)$ for a spatially distributed release of \acp{SM} can be obtained as follows
{\setlength{\abovedisplayskip}{1pt}
 \setlength{\belowdisplayskip}{1pt}
\begin{align}\label{eq:concentration_solution}
    c(x,t) = \mathbf{e}^\intercal(x) \mathbf{\hat{c}}(t) = \frac{N_{\mathrm{P}}}{L} \mathbf{e}^\intercal(x) \mathrm{e}^{\mathbf{A} t} \boldsymbol{\phi} \;.
\end{align}}
\subsection{Comparison Between Analytical Solution and Particle-Based Simulation}\label{sec:analytical_vs_PBS}
With the \acp{PBS}, we validate the accuracy of the simplification step of reducing the model of the \ac{3D} pipe to a \ac{1D} model. Additionally, we validate that for $N = 100$, i.e., when the Fourier series in \Equation{eq:fourier_series} is truncated to $2N+1 = 201$ terms, $c(x,t)$ is well approximated.

\begin{figure}
    \centering
    \vspace{-0.15cm}
    \includegraphics[width=1\linewidth]{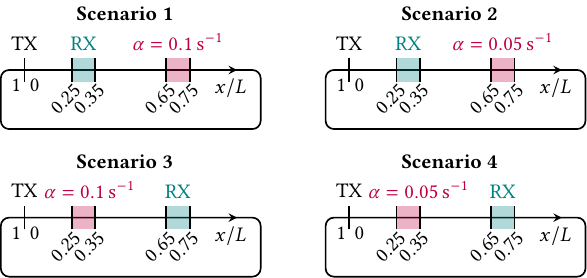}
    \vspace{-0.3cm}
    \caption{\small \ac{1D} representation of four different closed-loop scenarios considered for verification of the proposed model.}\Description{1D representation of four different closed-loop scenarios considered for verification of the proposed model.}
    \label{fig:3:evaluation scenarios}
    \vspace{-0.5cm}
\end{figure}

In particular, we validate the analytical solution in \eqref{eq:int_c} for an instantaneous point-release of \acp{SM}, cf. \Equation{eq:impulse_release_source}, and a system comprising two levels of damping, cf. \eqref{eq:ftwolevel}. For generality, we consider the four scenarios depicted in \Figure{fig:3:evaluation scenarios}, with two different damping constants $\alpha$ and two different positions of the  transparent \ac{RX} and damping region $[x_{\mathrm{a}}, x_{\mathrm{b}}]$, respectively. The values of $\alpha$ and the limits of the \ac{RX} and the damping region are shown in \Figure{fig:3:evaluation scenarios}. Outside the localized damping region, i.e., outside the interval $[x_{\mathrm{a}}, x_{\mathrm{b}}]$, a constant damping of $\beta = 0.01\, \si{\per\second}$ is considered.
The default values of all system parameters are given in \Table{Tab:Params} and are used throughout the paper, if not specified otherwise.

The analytical results in this section are based on \eqref{eq:c_impulse}, i.e., a point release. The received concentration follows as $c_{\mathrm{RX}}(t) = c(x_{\mathrm{RX}},t)$, i.e., by sampling $c(x,t)$ at axial position $x = x_{\mathrm{RX}}$. For \acp{RX} of spatial extent (as the ones in \Figure{fig:3:evaluation scenarios}), the received concentration $c_{\mathrm{RX}}(t)$ is obtained by integrating \eqref{eq:c_impulse} over the spatial interval $[x_{\mathrm{RX},\mathrm{a}},\, x_{\mathrm{RX},\mathrm{b}}]$
{\setlength{\abovedisplayskip}{1pt}
 \setlength{\belowdisplayskip}{1pt}
\begin{align}
    c_{\mathrm{RX}}(t) = \int_{x_{\mathrm{RX},\mathrm{a}}}^{x_{\mathrm{RX},\mathrm{b}}} c(x,t) \, \mathrm{d}x \;,
\end{align}}
where $[x_{\mathrm{RX},\mathrm{a}},\, x_{\mathrm{RX},\mathrm{b}}]$ with $0 \leq x_{\mathrm{RX},\mathrm{a}} < x_{\mathrm{RX},\mathrm{b}} \leq L$ denote the limits of the \ac{RX} region. For the latter, the center position of the \ac{RX} is given by $x_{\mathrm{RX}} = (x_{\mathrm{RX},\mathrm{a}} + x_{\mathrm{RX},\mathrm{b}})/2$.

\subsubsection{Details of the Particle-Based Simulation}
The \acp{PBS} are performed for a \ac{3D} straight pipe of length $L$ with  \textit{reflective radial boundaries} and \textit{periodic boundaries} at $x=0$ and $x=L$.
At time $t = 0$, $N_{\mathrm{P}}$ particles are uniformly distributed over the cross section of the pipe at $x=0$. For $t>0$, the particles in the pipe are propagated by diffusion and laminar flow in $x$ direction with a parabolic flow profile, i.e., $ v_x(r_{\mathrm{p}}) = 2v_{\mathrm{eff}}\left( 1 - \left(r_{\mathrm{p}}/r_0\right)^2\right)$. Here, $v_x$ and $r_{\mathrm{p}}$ are the particle's velocity in $x$ direction induced by flow and the distance of the particle from the $x$ axis, i.e., the centerline of the \ac{3D} pipe, respectively. The clearance of the \acp{SM} is realized as a first-order degradation reaction, implemented by the stochastic simulation algorithm (SSA)~\cite{Erban_Chapman_2020}. All results shown are normalized by $N_{\mathrm{P}}$.
\begin{table}[ht]
\centering
\vspace*{-0.3cm}
\caption{\small Parameters used in the \ac{PBS} and analytical solution.}
\vspace*{-0.3cm}
\label{Tab:Params}
\begin{tabular}{p{1.4cm} p{3.7cm} c l}
\hline
\textbf{Parameter} & \textbf{Description} & \textbf{Value} & \textbf{Unit} \\
\hline
\\[-2.4ex] 
$D_\mathrm{eff}$ & Effective diffusion coefficient & $5 \times 10^{-3}$ & $\si{\square\meter\per\second}$ \\
$L$ & Length & $6$ & $\si{\meter}$ \\
$r_{0}$ & Radius of pipe & $0.02$ & $\si{\meter}$\\
$v_\mathrm{eff}$ & Effective velocity & $0.1$ & $\si{\meter\per\second}$ \\
$\alpha$ & Degradation rate in $[x_{\mathrm{a}}, x_{\mathrm{b}}]$ & $\{0.05,0.1\}$ & $\si{\per\second}$ \\
$\beta$ & Degradation rate outside $[x_{\mathrm{a}}, x_{\mathrm{b}}]$ & $0.01$ & $\si{\per\second}$ \\
$N_{\mathrm{P}}$ & Number of particles & $10,000$ & -- \\
$\Delta t$ & PBS time step & $10^{-6}$ & $\si{\second}$ \\
$T$ & Total simulation time & $100$ & $\si{\second}$ \\
\hline
\end{tabular}
\vspace*{-0.5cm}
\end{table}
\subsubsection{Model Verification}
\begin{figure}[ht]
    \centering
    \vspace{-0.3cm}
    \includegraphics[width=0.98\linewidth]{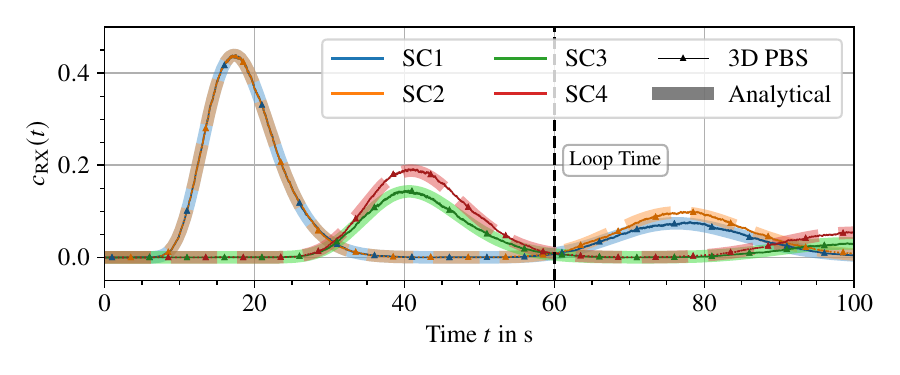}
    \vspace{-0.7cm}
    \caption{\small Received signal $c_\mathrm{RX}(t)$ obtained with the analytical solution \eqref{eq:c_impulse} (shaded lines) and by \ac{PBS} (solid lines). The different colors indicate the different scenarios shown in \Figure{fig:3:evaluation scenarios}.
    }\Description{Received signal $c_\mathrm{RX}(t)$ obtained with the analytical solution \eqref{eq:c_impulse} (shaded lines) and by \ac{PBS} (solid lines). The different colors indicate the different scenarios shown in \Figure{fig:3:evaluation scenarios}.}
    \vspace{-0.5cm}
    \label{fig:PBSvaANA}
\end{figure}

Figure~\ref{fig:PBSvaANA} shows the analytical and \ac{PBS} results for all four scenarios summarized in \Figure{fig:3:evaluation scenarios}. We observe that the results obtained from \ac{PBS} (solid lines) and the proposed analytical solution (shaded lines) are in excellent agreement, confirming the accuracy of the derived model. Moreover, the effects introduced by the closed-loop system (repeated observation of the same \acp{SM}) and the different types of damping can be observed. 
First, we observe that the received signal amplitude is damped over time, where the damping is more pronounced for Scenarios 1 and 3, cf. \Figure{fig:3:evaluation scenarios}, as rate $\alpha$ is higher in these cases, thus yielding, as expected, a stronger damping.  
For Scenarios 1 and 2, cf. \Figure{fig:3:evaluation scenarios}, where the \ac{RX} is located between the \ac{TX} and the damping region, the first observed peak is affected by the global damping only, while subsequent peaks are attenuated by the effects of local damping constant $\alpha$ and global damping constant $\beta$. This observation suggests that damping at a properly chosen location and with a suitable rate $\alpha$ can mitigate \ac{ISI} in an \ac{MC} system without large effects on the first peak of the transmission (see Section~\ref{sec:isi}). 
In comparison, in Scenarios 3 and 4, cf. \Figure{fig:3:evaluation scenarios}, where the \ac{RX} is placed behind the damping region, already the first peak observed at the \ac{RX} is damped significantly. 
%
\section{Communication System and Analysis}\label{sec:isi}
In this section, we use the analytical model for the \ac{SM} concentration, $c(x, t)$, obtained in Section~\ref{sec:description_and_solution}, to investigate the impact of a closed-loop system with constant and/or localized damping on \ac{MC}-based information transmission.
In the following, we consider a closed-loop \ac{MC} system employing \ac{OOK} modulation. In this setup, the \ac{TX} releases $N_{\mathrm{P}}$ signaling molecules for the transmission of bit $1$ and does not release any molecules for bit $0$. The binary bit sequence can be represented by an impulse train as follows
{\setlength{\abovedisplayskip}{1pt}
 \setlength{\belowdisplayskip}{1pt}
\begin{equation}
    b(t) = \sum_{p=0}^{B-1} \tilde{b}_p \delta(t - p T_{\mathrm{S}} - t_0)\;, \label{eq:sequence}
\end{equation}}
where $B$, $\tilde{b}_p$, $T_{\mathrm{S}}$, and $t_0$ denote the total number of transmitted bits, i.e., the sequence length, the $p$-th transmitted bit with $p \in \{0, 1, \dots, B-1\}$ and $\tilde{b}_p = 1$ for bit $1$ and $\tilde{b} = 0$ for bit $0$, the symbol interval duration, and the transmission start time, respectively. 
Upon release of the binary sequence in \eqref{eq:sequence} by the \ac{TX}, the received signal $r(t)$ at the \ac{RX} results from the linear superposition of the individual system responses $c(x,t)$, cf. \Equation{eq:concentration_solution}, caused by different symbols. Consequently, the received signal $r(t)$ is obtained by convolving $c_{\mathrm{RX}}(t)$ with $b(t)$ as follows
{\setlength{\abovedisplayskip}{1pt}
 \setlength{\belowdisplayskip}{1pt}
\begin{equation}\label{eq:receivedSig}
    r(t) = b(t) \ast c_{\mathrm{RX}}(t)\;,
\end{equation}}
where $\ast$ denotes the convolution operator. For simplicity, all results for $r(t)$ presented in \Section{sec:evaluation} are normalized by $N_{\mathrm{P}}$. Additionally, we assume that $\tilde{b}_p = 1$ and $\tilde{b}_p = 0$ occur equally often in the binary sequence. In the next section, the signal $c(x, t)$ is used to classify different types of \ac{ISI}, before the analysis is extended to the received signal $r(t)$ in \Section{sec:evaluation}.
\subsection{Inter-Symbol Interference}
A key challenge in \ac{MC} is the occurrence of \ac{ISI}. Specifically, the modulated signal can overlap with signals from previous and subsequent symbol intervals, leading to causal and anticausal \ac{ISI}~\cite{wang2020understanding} at the \ac{RX}, cf. \Figure{fig:2:system model}. 
In most advection-diffusion-based \ac{MC} systems considered so far, any released \ac{SM} appears only once at the \ac{RX}. However, in closed-loop systems, \acp{SM} persist as they circulate repeatedly inside the system, causing \ac{ISI} multiple times. In other words, a closed-loop system represents a multi-path propagation environment as each circulation loop delivers the \acp{SM} to the \ac{RX} again, while only the first arrival contains the desired\footnote{Of course, this statement only holds when symbol-by-symbol detection is used.} information. While the length of the direct path is given by the distance between \ac{TX} and \ac{RX}, each subsequent path is elongated by the loop length $L$. As a result, these 'looped' versions of the signal arrive as different dispersed versions of the original signal at staggered times at the \ac{RX}, interfering with both neighboring symbols and \textit{those transmitted much later}.

The resulting unique types of \ac{ISI} in closed-loop systems, introduced as \textit{inter-loop} \ac{ISI} and \textit{offset} \ac{ISI} in~\cite{scherer2025closed}, require targeted mitigation strategies, such as customized localized \ac{SM} clearance blocks or tailor-made equalization techniques. Developing such strategies is crucial, yet challenging, as these types of \ac{ISI} have not been quantitatively characterized to date. Therefore, before mitigation strategies can be developed, it is imperative to provide a clear mathematical definition and characterization of each type of \ac{ISI} occurring in closed-loop systems.
Assuming that \ac{TX} and \ac{RX} in a closed-loop \ac{MC} system are positioned such that the first and highest peak of the received signal is due to \acp{SM} propagating in the direction of the flow, the time of arrival of the first peak at the \ac{RX} is given by \cite[Eq. (14)]{Wicke2018}
{\setlength{\abovedisplayskip}{1pt}
 \setlength{\belowdisplayskip}{1pt}
\begin{align}
    t_{\mathrm{p}} = t_0 + \frac{-D_{\mathrm{eff}} + \sqrt{D_{\mathrm{eff}}^2 + (x_{\mathrm{RX}} - x_{\mathrm{TX}})^2 v_\mathrm{eff}^2}}{v_\mathrm{eff}^2}\;,\label{eq:peak_time}
\end{align}}
where $x_{\mathrm{TX}} = x_{\mathrm{w}}/2$ for spatial distributed \ac{SM} release (see Section~\ref{sssec:spatialRelease})\footnote{We note that for special cases, such as diffusion-dominated propagation of \acp{SM} against the flow direction, the peak time $t_{\mathrm{p}}$ in~\Equation{eq:peak_time}, which assumes downstream communication, is not valid. However, as such scenarios are considered to have limited practical relevance, they are not addressed in this work.}. In the following, we discuss the different types of \ac{ISI} occurring in closed-loop systems in more detail.

\subsubsection{Channel ISI}

Channel \ac{ISI} arises due to the overlap of \acp{SM} from neighboring symbol intervals during the propagation from \ac{TX} to \ac{RX}, and has been extensively studied in the \ac{MC} literature \cite{Pierobon_ISI}. Specifically, channel \ac{ISI} is caused by \acp{SM} arriving at the \ac{RX} outside of their designated symbol interval of duration $T_{\mathrm{S}}$ centered around the peak arrival time $t_{\mathrm{p}}$, without having yet completed a full circulation in the closed-loop system.

To obtain a mathematical description for the channel \ac{ISI} in closed-loop systems, only the contribution of those \acp{SM} that cause interference on their direct path from \ac{TX} to \ac{RX} should be included. The contribution of the direct path $c^{\mathrm{open}}_{\mathrm{RX}}(t)$ to the received signal $c_{\mathrm{RX}}(t)$ can be obtained either by solving the corresponding open-loop system, i.e., by considering \eqref{eq:PDE_general} without periodic boundary conditions, as shown for a closed-loop system without damping in \cite[Fig.~8]{Schaefer2024ChorioallantoicMembraneModel} or by taking the limit value for $L\to\infty$ of the closed-loop model solution in \eqref{eq:int_c} derived in this work. 
Hence, the concentration component $c_{\mathrm{c}}(t)$ attributable to channel \ac{ISI} can be defined as
{\setlength{\abovedisplayskip}{1pt}
 \setlength{\belowdisplayskip}{1pt}
\begin{equation}
    c_{\mathrm{c}} (t) \!= \!
    \begin{cases} 
    c^{\mathrm{open}}_{\mathrm{RX}}(t) & \!\!\!\forall t \in \left[t_0,\max(t_0, t_{\mathrm{p}}-\frac{T_{\mathrm{S}}}{2})\right)\! \cup \!\left[t_{\mathrm{p}}+\frac{T_{\mathrm{S}}}{2}, \infty\right) \;, \\ 
    0 & \!\!\text{else} \;.
    \end{cases}
\end{equation}}
The desired signal component $c_{\mathrm{d}} (t)$ follows as
{\setlength{\abovedisplayskip}{1pt}
 \setlength{\belowdisplayskip}{1pt}
\begin{equation}
    c_{\mathrm{d}} (t) = 
    \begin{cases} 
    c^{\mathrm{open}}_{\mathrm{RX}}(t) & \forall t \in \left[\max(t_0, t_{\mathrm{p}}-\frac{T_{\mathrm{S}}}{2}), t_{\mathrm{p}}+\frac{T_{\mathrm{S}}}{2}\right) \;, \\ 
    0 & \text{else} \;,
    \end{cases}
\end{equation}}
i.e., the signal centered around $t_{\mathrm{p}}$.
\subsubsection{Inter-Loop ISI}
The second type of \ac{ISI} in closed-loop systems is \textit{inter-loop \ac{ISI}}. It emerges due to the recirculation of \acp{SM} in the system, i.e., after completion of at least one circulation, \acp{SM} reappear at the \ac{RX} as peaks (cf. \Figure{fig:PBSvaANA}) and possibly interfere with \acp{SM} from subsequent symbol intervals. 
To mathematically quantify inter-loop \ac{ISI}, we can subtract the open-loop solution $c^{\mathrm{open}}_{\mathrm{RX}}(t)$ (representing the contribution from the direct path) from the total concentration observed in the closed-loop system $c_{\mathrm{RX}}(t)$ as follows
{\setlength{\abovedisplayskip}{1pt}
 \setlength{\belowdisplayskip}{1pt}
\begin{equation}
    c_{\mathrm{i}}(t) = \begin{cases}
        c_{\mathrm{RX}}(t) - c^{\mathrm{open}}_{\mathrm{RX}}(t) & \forall t \in [t_0, t_{\mathrm{i}}] \;, \\
        0 & \text{else} \;,
    \end{cases}
\end{equation}}
where $t_{\mathrm{i}}$ denotes the time when inter-loop \ac{ISI} transitions into a constant signal, which we refer to as an offset. This time can be identified based on the fraction of the global signal decay, described by the $0$th mode of the respective solution, and the received signal as follows
{\setlength{\abovedisplayskip}{1pt}
 \setlength{\belowdisplayskip}{1pt}
\begin{align}\label{eq:transTime}
    t_{\mathrm{i}} := \inf \left\{ t \in \mathbb{R}_{\geq 0} \;\middle|\; 
    \forall \tau \geq t:
    \left| 
    \frac{\hat{c}_0(\tau) - \hat{c}^{\mathrm{open}}_{0}(\tau)}{c_{\mathrm{RX}}(\tau) - c^{\mathrm{open}}_{\mathrm{RX}}(\tau) }
    \right| \geq \epsilon
    \right\}\;,
\end{align}}
where $\hat{c}_0(t)$ denotes the $n = 0$ mode of $c(x,t)$ in \eqref{eq:int_c}, similarly $\hat{c}^{\mathrm{open}}_{0}(t)$ denotes the $0$th mode of the corresponding open-loop solution. Therefore, \eqref{eq:transTime} ensures that as soon as the zero-frequency component, i.e., the offset component (numerator in \eqref{eq:transTime}), exceeds a threshold fraction $\epsilon$ of the total received inter-loop \ac{ISI} signal (denominator in \eqref{eq:transTime}), the received signal is considered as offset \ac{ISI}, which is explained next.
\subsubsection{Offset ISI}
The third type of \ac{ISI} occurring in closed-loop systems is offset \ac{ISI}. It represents the persistent shift in signal baseline caused by the accumulation of \acp{SM} circulating in the system. As \acp{SM} disperse during circulation, their influence on the signal received at the \ac{RX} transitions from discrete interferences, denoted as inter-loop \ac{ISI}, to a continuous offset in the received signal. The offset \ac{ISI} part of the signal can be defined as follows
{\setlength{\abovedisplayskip}{1pt}
 \setlength{\belowdisplayskip}{1pt}
\begin{equation}
    c_{\mathrm{o}}(t) = \begin{cases} 
    c_{\mathrm{RX}}(t) - c^{\mathrm{open}}_{\mathrm{RX}}(t) & \forall t > t_{\mathrm{i}} \;, \\
    0 & \text{else} \;.
    \end{cases}\label{eq:offset_isi}
\end{equation}}
\subsection{Closed-Loop Equilibrium Concentration}
As an additional quantity for the characterization of closed-loop systems, we define an equilibrium concentration of \acp{SM} in the system, $r_{\mathrm{eq}}$, as the concentration where, on average per symbol interval, the number of \acp{SM} removed from the system due to damping equals the number of \acp{SM} released into the system, as follows
{\setlength{\abovedisplayskip}{1pt}
 \setlength{\belowdisplayskip}{1pt}
\begin{align}
    &\underbrace{N_{\mathrm{P}}/2}_{\raisebox{4pt}{\scriptsize\textnormal{average release per symbol}}}
    =
    \underbrace{
        r_{\mathrm{eq}} T_{\mathrm{S}} \big((x_{\mathrm{b}}{-}x_{\mathrm{a}}) \alpha + (L{-}(x_{\mathrm{b}}{-}x_{\mathrm{a}})) \beta \big)
    }_{\raisebox{4pt}{\scriptsize\textnormal{average damping per symbol}}}
    \nonumber\\[-0.2cm]
    & \rightarrow
    r_{\mathrm{eq}} =
    \frac{N_{\mathrm{P}}}{2 T_{\mathrm{S}} \big((x_{\mathrm{b}}{-}x_{\mathrm{a}}) \alpha + (L{-}(x_{\mathrm{b}}{-}x_{\mathrm{a}})) \beta \big)}
    \;.\label{eq:equilibrium_concentration}
\end{align}}
This equilibrium concentration inherently captures the accumulation of \acp{SM} within the closed-loop system after repeated \ac{SM} injections. This provides an estimate of the operational concentration range of \acp{SM} over extended transmission times.
%
\section{Evaluation}\label{sec:evaluation}
In this section, we analyze the various types of \ac{ISI} occurring in closed-loop systems for long transmission sequences and investigate the effect of localized damping on the received signal. 
\subsection{Temporal Evolution of ISI During Continuous Transmission}
\begin{figure*}[!tbp]
\centering 
    \vspace{-0.35cm}
  \includegraphics[width = 0.98\textwidth]{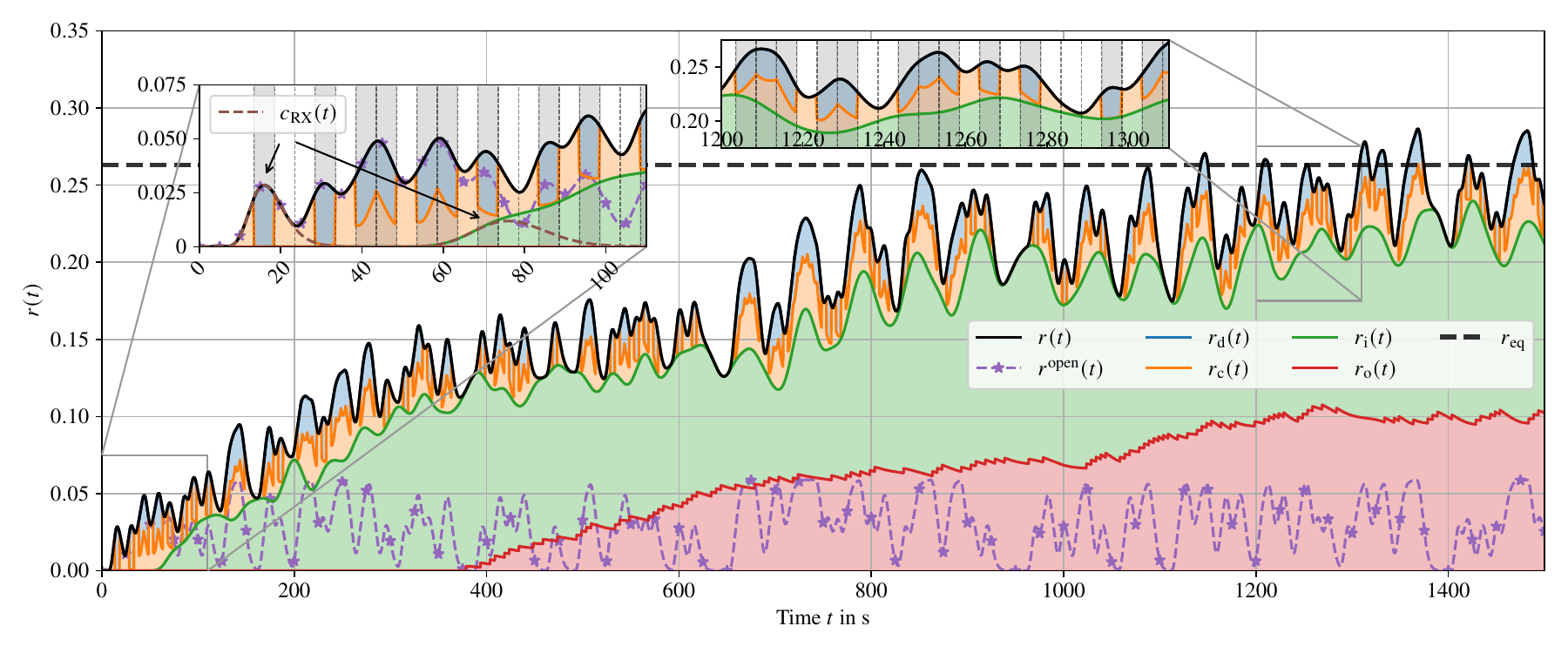}\vspace*{-7mm}
  \caption{\small$r(t)$ is decomposed into the desired received signal $r_{\mathrm{d}}(t)$, channel \ac{ISI} $r_{\mathrm{c}}(t)$, inter-loop \ac{ISI} $r_{\mathrm{i}}(t)$, and offset \ac{ISI} $r_{\mathrm{o}}(t)$. Additionally, $r^{\mathrm{open}}(t)$ and the received signal due to the first transmitted bit 1 ($c_{\mathrm{RX}}(t)$, shown in the upper-left inset) are provided. Here, $x_{\mathrm{RX}} = 0.3 L$, $N = 1000$, $\epsilon = 0.8$, $\alpha = 0.01\,\si{\per\second}$, $\beta = 0.001\,\si{\per\second}$, and a rectangular \ac{TX} with $x_{\mathrm{w}} = 0.3 \, \si{\meter}$, cf. \Section{sssec:spatialRelease}, are used. The shaded areas in the insets indicate the transmission of bit 1.}\Description{$r(t)$ is decomposed into the desired received signal $r_{\mathrm{d}}(t)$, channel \ac{ISI} $r_{\mathrm{c}}(t)$, inter-loop \ac{ISI} $r_{\mathrm{i}}(t)$, and offset \ac{ISI} $r_{\mathrm{o}}(t)$. Additionally, $r^{\mathrm{open}}(t)$ and the received signal due to the first transmitted bit 1 ($c_{\mathrm{RX}}(t)$, shown in the upper-left inset) are provided. Here, $x_{\mathrm{RX}} = 0.3 L$, $N = 1000$, $\epsilon = 0.8$, $\alpha = 0.01\,\si{\per\second}$, $\beta = 0.001\,\si{\per\second}$, and a rectangular \ac{TX} with $x_{\mathrm{w}} = 0.3 \, \si{\meter}$, cf. \Section{sssec:spatialRelease}, are used. The shaded areas in the insets indicate the transmission of bit 1.}
  \label{fig:ISI_Forms}
  \vspace*{-0.45cm}
\end{figure*}
First, we investigate the emergence and progression of the different types of \ac{ISI} during continuous information transmission. To analyze the system behavior, we apply the formal definitions of the different types of \ac{ISI} from \Section{sec:isi} on a symbol-by-symbol basis and compute the received signal $r(t)$ by convolving the input sequence with the system response, see \eqref{eq:receivedSig}. As evaluation scenario, we consider Scenario~1, as depicted in \Figure{fig:3:evaluation scenarios}.

In this scenario, the \ac{TX} transmits a binary sequence $b(t)$, as defined in \eqref{eq:sequence}, of length $B = 300$ bits. As symbol duration, $T_{\mathrm{S}} = 5 \,\si{\second}$ is used. The corresponding received signal $r(t)$ (solid black line) along with its constituent components is presented in \Figure{fig:ISI_Forms}. Specifically, $r(t)$ consists of the desired signal $r_{\mathrm{d}}(t)$ (blue), channel \ac{ISI} $r_{\mathrm{c}}(t)$ (orange), inter-loop \ac{ISI} $r_{\mathrm{i}}(t)$ (green), and offset \ac{ISI} $r_{\mathrm{o}}(t)$ (red). 
For reference, the contribution of the first transmitted bit $1$ is shown separately as $c_{\mathrm{RX}}(t)$ (dashed brown line in the upper-left inset), along with the received signal of a corresponding open-loop system, $r^{\mathrm{open}}(t)$, and the system’s equilibrium concentration, $r_{\mathrm{eq}}$ (horizontal dotted line), cf. \Equation{eq:equilibrium_concentration}.

From the insets in \Figure{fig:ISI_Forms}, we observe that $r(t)$ increases for the transmission of bit $1$s. The upper left shows that during the first $9$ bits, $r(t)$ comprises only the desired signal and channel \ac{ISI}, as no \acp{SM} have yet completed a full loop through the system. 
The emergence of the second peak from the first bit $1$, after one circulation, shown in the upper-left inset, marks the onset of inter-loop \ac{ISI}. Thereafter, inter-loop \ac{ISI} is progressively increasing, as the cumulative contributions of \acp{SM} completing one or more full loops increases. 
At around $t = 450\,\si{\second}$, the contribution from the first transmitted bit becomes sufficiently damped to satisfy the criteria for offset \ac{ISI}, cf. \Equations{eq:transTime}{eq:offset_isi}, which thereafter increases, before it saturates at around $t = 1300\,\si{\second}$.

Towards the end of the transmission sequence, we observe from \Figure{fig:ISI_Forms} that $r(t)$ approaches the equilibrium concentration $r_{\mathrm{eq}}$. We note that in this phase of the transmission, the peak amplitude of individual bit-1 transmissions received at the \ac{RX} is approximately $0.025$, which is nearly one order of magnitude below $r_{\mathrm{eq}} = 0.26$.
Given this ratio of $0.025$ from \acp{SM} compared to $0.26$ from residual molecules originating from previous transmissions, reliable communication is expected to become increasingly challenging, if not impossible. Note that, while we show the concentration here, i.e., the expected received signal, the actual received signal will be noisy due to diffusion and receiver noise and the desired signal $r_{\mathrm{d}}(t)$ may be completely obscured.

These observations highlight the importance of understanding the distinct \ac{ISI} components and the role of the equilibrium concentration $r_{\mathrm{eq}}$ when analyzing or designing \ac{MC} systems for long-term operation in closed-loop environments. Specifically, systems with high $r_{\mathrm{eq}}$ values compared to the desired signal amplitudes require effective \ac{ISI} mitigation strategies to enhance communication reliability. One such strategy involves increased localized damping to suppress the endless circulation of \acp{SM}, as explored in the following section.

\subsection{Effect of Localized Damping on Emulating Open-Loop Behavior}

\begin{figure}[!tbp]
\centering 
    \vspace{-0.1cm}
  \includegraphics[width = 1\columnwidth]{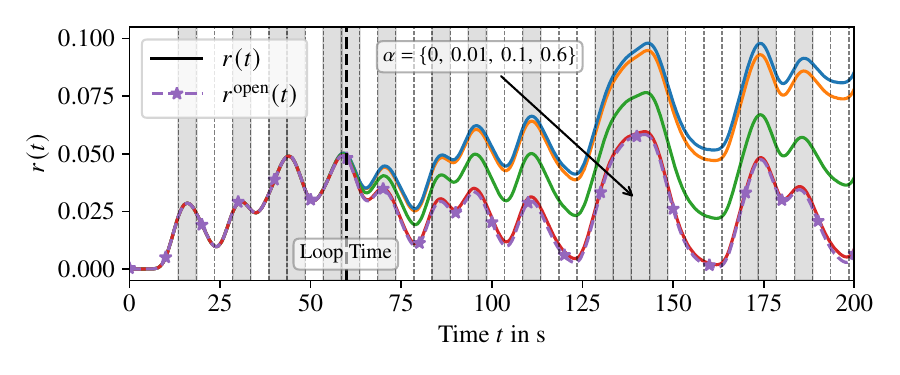}\vspace*{-6mm}
  \caption{\small Evaluation of the evolution of $r(t)$ for different localized damping constants $\alpha = \{0, 0.01, 0.1, 0.6\}\,\si{\per\second}$. Additionally, for comparison, $r^{\mathrm{open}}(t)$ is shown. Here, $x_{\mathrm{RX}} = 0.3 L$, $N=250$, $\beta = 0.001\,\si{\per\second}$, and a rectangular \ac{TX} with $x_{\mathrm{w}} = 0.3 \, \si{\meter}$, cf. \Section{sssec:spatialRelease}, are used. Shaded areas correspond to the transmission of bit 1.}\Description{Evaluation of the evolution of $r(t)$ for different localized damping constants $\alpha = \{0, 0.01, 0.1, 0.6\}\,\si{\per\second}$. Additionally, for comparison, $r^{\mathrm{open}}(t)$ is shown. Here, $x_{\mathrm{RX}} = 0.3 L$, $N=250$, $\beta = 0.001\,\si{\per\second}$, and a rectangular \ac{TX} with $x_{\mathrm{w}} = 0.3 \, \si{\meter}$, cf. \Section{sssec:spatialRelease}, are used. Shaded areas correspond to the transmission of bit 1.}
  \label{fig:local_attenuation}
  \vspace*{-0.5cm}
\end{figure}

In \Figure{fig:local_attenuation}, we investigate the influence of localized damping on the received signal $r(t)$. In particular, we consider the setup from Scenario~1 in \Figure{fig:3:evaluation scenarios}, and vary the damping constant from $\alpha = 0\,\si{\per\second}$ (blue) to $\alpha = 0.6\,\si{\per\second}$ (red). For comparison, \Figure{fig:local_attenuation} also includes the received signal $r^{\mathrm{open}}(t)$ (dotted) of the corresponding open-loop system.

From \Figure{fig:local_attenuation}, we observe that until $t = 60\,\si{\second} $ the received signals $r(t)$ for all considered damping constant $\alpha$ coincide, as the damping region is positioned downstream of the \ac{RX} (see \Figure{fig:3:evaluation scenarios}). Consequently, the damping only affects \acp{SM} revisiting the \ac{RX} after one circulation. Therefore, in the considered arrangement, the damping only affects, and could possibly mitigate, inter-loop and offset \ac{ISI}, but not channel \ac{ISI}. As shown in \Figure{fig:local_attenuation}, the received signal $r(t)$ is attenuated to different degrees depending on the value of $\alpha$. This occurs because the portion of the signal originating from \acp{SM} that have completed at least one cycle is attenuated. Moreover, \Figure{fig:local_attenuation} shows that when $\alpha$ is sufficiently high (e.g., $\alpha = 0.6\,\si{\per\second}$), the received signal $r(t)$ closely matches the corresponding open-loop system's $r^{\mathrm{open}}(t)$. This suggests that the effects of interference caused by the specific properties of closed-loop systems can be mitigated by localized damping. This finding is particularly important for designing practical \ac{MC} systems, as it may allow closed-loop systems to operate in a regime where analytical tools and models already developed for open-loop \ac{MC} systems can be applied directly.
%
\section{Conclusion}\label{sec:conclusion}
In this paper, a novel physics-based model for the propagation of molecular signals in closed-loop \ac{MC} systems was proposed along with a reformulation of the system's descriptive equations that allows for efficient numerical evaluation.
The model encompasses arbitrary spatio-temporal molecule release profiles and can represent different types of molecule degradation that are practically relevant for natural and synthetic \ac{SM} degradation/clearance.
The presented results allow, for the first time, to rigorously distinguish different types of \ac{ISI} that can occur in closed-loop \ac{MC} systems.
Furthermore, they reveal under which conditions the open loop assumption made in many \ac{MC} models is an accurate approximation of a closed-loop system.

Extending the model proposed in this paper to branched vessel topologies can be an interesting topic for future research.

%


\bibliographystyle{ACM-Reference-Format}
\bibliography{literature.bib}
\end{document}